\documentclass[a4paper,11pt]{article}
\usepackage{pos}

\usepackage[capitalise]{cleveref}
\usepackage{physics}
\usepackage{comment}
\usepackage{xcolor}
\colorlet{linkequation}{blue}
\usepackage{subcaption}
\usepackage{wrapfig}
\usepackage{comment}
\usepackage[capitalise]{cleveref}

\let\OLDthebibliography\thebibliography
\renewcommand\thebibliography[1]{
  \OLDthebibliography{#1}
  \setlength{\parskip}{0pt}
  \setlength{\itemsep}{0pt plus 0.3ex}
}

\newcommand{\x}{\boldsymbol{x}}
\newcommand{\p}{\boldsymbol{p}}

\newcommand{\tauhydro}{\tau_\text{hydro}}

\newcommand{\wtilde}{\tilde{w}}
\newcommand{\eff}{\text{eff}}

\crefname{pluralequation}{eqs.}{eqs.}
\Crefname{pluralequation}{Eqs.}{Eqs.}

\crefname{section}{sec.}{sec.}
\Crefname{section}{Sec.}{Sec.}

\crefname{figure}{fig.}{fig.}
\Crefname{figure}{Fig.}{Fig.}

\crefname{table}{tab.}{tab.}
\Crefname{table}{Tab.}{Tab.}

\crefformat{equation}{\textup{\hypersetup{linkcolor=black,linkbordercolor=linkequation}#2eq.~({\color{blue}#1})#3}}
\Crefformat{equation}{\textup{\hypersetup{linkcolor=black,linkbordercolor=linkequation}#2Eq.~({\color{blue}#1})#3}}
\Crefformat{pluralequation}{\textup{\hypersetup{linkcolor=black,linkbordercolor=linkequation}#2Eqs.~({\color{blue}#1})#3}}

\crefname{chapter}{sec.}{sec.}
\Crefname{chapter}{Sec.}{Sec.}

\crefname{secinapp}{app.}{app.}
\Crefname{secinapp}{App.}{App.}

\title{Pre-equilibrium photon production in QCD Kinetic Theory}

\author[a]{Oscar Garcia-Montero}
\author[b]{Aleksas Mazeliauskas}
\author*[a]{Philip Plaschke}
\author[a]{Sören Schlichting}

\affiliation[a]{Fakult\"at f\"ur Physik, Universit\"at Bielefeld\\
D-33615 Bielefeld, Germany}

\affiliation[b]{Institut f\"ur Theoretische Physik, Universit\"at Heidelberg \\
69120 Heidelberg, Germany}

\emailAdd{pplaschke@physik.uni-bielefeld.de}

\notes{\note{\textbf{Acknowledgment}: We thank Stephan Ochsenfeld and Xiaojian Du for their valuable discussions. OGM, PP and SS acknowledge support by the Deutsche Forschungsgemeinschaft (DFG, German Research Foundation) through the CRC-TR 211 ‘Strong-interaction matter under extreme conditions’-project number 315477589 – TRR 211. The work of AM is funded by the DFG – project number 496831614. OGM and SS acknowledge also support by the German Bundesministerium
für Bildung und Forschung (BMBF) through Grant No. 05P21PBCAA. The authors acknowledge computing time provided by the Paderborn Center for Parallel Computing (PC2) and the National Energy Research Scientific Computing Center, a DOE Office of Science User Facility supported by the Office of Science of the U.S. Department of Energy under Contract No. DE-AC02-05CH11231.}}

\abstract{We use QCD kinetic theory to compute photon production in the chemically equilibrating Quark-Gluon Plasma created in the early stages of high-energy heavy-ion collisions. We show that the photon spectrum radiated from an attractor evolution satisfies a simple scaling form in terms of the specific shear viscosity $\eta/s$ and entropy density $dS/d\zeta \sim {\scriptstyle \left(T\tau^{1/3}\right)^{3/2}}_\infty$. We confirm the analytical predictions with numerical kinetic theory simulations. We use the extracted scaling function to compute the pre-equilibrium photon contribution in $\sqrt{s_{NN}}=2.76\,\text{TeV}$ 0-20\% PbPb collisions. We demonstrate that our matching procedure allows for a smooth switching from pre-equilibrium kinetic to thermal hydrodynamic photon production.}

\FullConference{HardProbes2023\\
 26-31 March 2023\\
 Aschaffenburg, Germany\\}

\begin{document}
\maketitle

\section{Introduction}
High-energy heavy-ion collisions (HICs) produce a hot and dense Quark-Gluon plasma (QGP), whose dynamic is described by Quantum Chromo Dynamics (QCD). Initially, the QGP is far from equilibrium and undergoes a stage of chemical and kinetic equilibration before reaching the thermalization described by hydrodynamics \cite{Schlichting:2019abc,Berges:2020fwq}. During all of these stages, as well as during the hadronic stage later on, electromagnetic probes such as photons and dileptons are produced. Once they are produced, they do not interact with the medium anymore making them a powerful tool to probe the various stages of HICs. Photons from hard scatterings at the inital stages of the collision (prompt photons) as well as photons from hadronic decays are relatively well understood. In terms of in-medium radiation, the pre-equilbirium contribution is, by now, subject to detailed studies. Nevertheless, up to now, in most studies the pre-equilibrium photon production was neglected or described in a simplified way. Although there are notable exceptions, a full description of pre-equilibrium photon production was still missing.

We compute the pre-equilibrium photon spectrum using leading order QCD kinetic theory. Our calculation includes all leading-order processes for quarks, gluons and photons. Elastic in-medium scattering matrix elements are modified using an isotropic screening regulator. Regarding the inelastic collisions, the Landau-Pomeranchuk-Midgdal (LMP) effect for in-medium splitting rates is treated via an effective vertex resummation in the formalism of AMY \cite{Arnold:2002zm}. For photon production in particular we include elastic pair annihilation and Compton scattering (elastic processes) as well as inelastic pair annihilation and Bremsstrahlung (inelastic processes) into our calculations.

\subsection{Time integrated spectrum}
In earlier works the thermalization of the QGP in QCD kinetic theory has been studied in much detail (see e.g. \cite{Du:2020dvp} and refs. therein). One important feature that was found is the emergence of hydrodynamic attractors. It was shown, that characteristics of the QGP, e.g. energy or pressure anisotropy as a function of the scaled time variable $\tilde{w}=\tau T/(4\pi \eta/s)$, follow a universal scaling function. The scaling in terms of $\wtilde$ was also found in the chemical equilibration of the QGP and was recently used to estimate dilepton production from pre-equilibrium stages \cite{Coquet:2021gms,Coquet:2021lca}. Therefore it is reasonable to assume that the radiation of photons off equilibrating fermions might will also follow a similar scaling. In this section we will derive such a universal scaling formula for photons produced in the pre-equilibrium phase of the QGP for a boost-invariant and transversely homogeneous expansion.

\subsection{Universal scaling of pre-equilibrium photon spectrum}
Integrated out space-time rapidity, the photon spectrum can be written as
\begin{align}\label{eq:DefaultProductionRate}
    \frac{dN_\gamma}{d^2\x_T d^2\p_T dy} = \frac{\nu_{\gamma}}{(2\pi)^3}\int_{\tau_\text{min}}^{\tau_\text{max}} d\tau \tau \int d^3\tilde{p} C_\gamma[f](X,\tilde{\Vec{{p}}}) \delta^{(2)} \pqty{\tilde{\p}_T - \p_T} \, ,
\end{align}
where we combined all four processes contributing at leading order into a single collision integral. By scaling the momentum with the effective temperature obtained from Landau-matching conditions as well as assuming that the system follows a non-equilibirium energy attractor, a lengthy calculation shows that the integral over the Bjorken time $\tau$ can be rewritten according to

\begin{align} \label{eq:ScaledSpectrum}
    \frac{dN_\gamma}{d^2\x_T d^2\p_T dy} &= \frac{(4\pi\eta/s)^2\nu_{\gamma}}{(2\pi)^3} \int_{\wtilde_\text{min}}^{\Tilde{w}_{\text{max}}} d\Tilde{w} ~\mathcal{F}(\wtilde) \\
    & \qquad \times \int d^3 {\displaystyle \left(\frac{\tilde{p}}{T_{\text{eff}}(\tau)} \right)} ~\Bar{C}_\gamma \pqty{\wtilde, \displaystyle \frac{\Vec{p}}{T_\eff(\tau)}} \delta^{(2)} \left( \frac{\mathcal{E}^{3/8}(\Tilde{w})}{\Tilde{w}^{1/2}} ~\frac{\tilde{\p}_T}{T_{\text{eff}}(\tau)} - \sqrt{4\pi} ~\frac{\sqrt{\eta/s}\p_T}{\left( T\tau^{1/3} \right)_\infty^{3/2}} \right) \nonumber \, ,
\end{align}
where we defined $\mathcal{F}(\wtilde) \equiv \frac{\mathcal{E}^{3/4}(\Tilde{w})}{\frac{3}{4} - \frac{1}{4} \mathcal{P}(\Tilde{w})}$ with $\mathcal{E}$ being the energy attractor \cite{Giacalone:2019ldn} and $\mathcal{P} = p_L/e$. The quantity ${\scriptstyle \left( T\tau^{1/3} \right)_\infty^{3/2}}$ is related to the entropy density per unit rapidity ${\scriptstyle \left( T\tau^{1/3} \right)_\infty^{3/2}} \sim (\tau s)^{1/2}$, while $\bar{C}_\gamma$ is a dimensionless function s.t. $C_\gamma(\tau,\Vec{p}) = T_\eff(\tau) \Bar{C}_\gamma \pqty{\wtilde, \displaystyle \Vec{p}/T_\eff(\tau)}$.

We note that the photon production rate $\Bar{C}_\gamma$ itself also explicitly contains the strong coupling constant. However, by looking at the ratio of the non-equilibrium rate to the thermal rate, this additional coupling dependence can be canceled out. By this logic we define the universal scaling function
\begin{align}\label{eq:UniversalScaling}
     \mathcal{N}_{\gamma}(\tilde{w},\bar{\mathbf{p}}_T) \equiv \frac{1}{(\eta/s)^2 ~\tilde{C}_\gamma^\text{ideal}} \frac{dN_\gamma}{d^2\x_T d^2\p_T dy} \, ,
\end{align}
such that the spectrum is naturally written in terms of the scaled transverse momentum $\bar{\mathbf{p}}_T = {(\eta/s)^{1/2}} \p_T/{\scriptstyle \left(\tau^{1/3} T \right)^{3/2}_\infty} = (\pi\nu_\text{eff}/90)^{1/2}(4\pi\eta/s)^{1/2} \p_T/(s T)_\infty^{1/2} $ and exhibits an overall $(\eta/s)^2$ scaling. The normalization $\tilde{C}_\gamma^\text{ideal}$ is obtained as a moment of the equilibrium photon rate (see next section).

\subsection{Thermal photon spectrum}
Starting from \Cref{eq:DefaultProductionRate} we can also derive the idealized thermal spectrum if we assume the system is in equilibrium for all times. In this case $T_\text{eff}= T(\tau) =(\tau^{1/3}T)_\infty/\tau^{1/3}$ and the collision integral only depends on $z=|\vec{p}|/T(\tau) = p_T\cosh \zeta/T(\tau)$. A detailed but straightforward calculation ends in
\begin{align}\label{eq:idealfinite}
     \frac{dN_\gamma}{d^2\x_T d^2\p_T dy} =& \frac{\nu_\gamma}{(2\pi)^3} \frac{(\eta/s)^2}{\bar{p}_T^4}4\bigg( \int_{\sqrt{4 \pi \tilde w_\text{min}} \bar{p}_T}^\infty \hspace{-0.8cm} dz  z^4 \,  \bar{C}_{\gamma}[f](z)  \bqty{1-{\textstyle \frac{4\pi \tilde{w}_\text{min} \bar{p}_T^2}{z^2}} }^{\frac{1}{2}} \bqty{1 + {\textstyle \frac{4 \pi \tilde{w}_\text{min} \bar{p}_T^2}{2 z^2 }}} \\
     &\qquad -\int_{\sqrt{4 \pi\tilde w_\text{max}} \bar{p}_T}^\infty \hspace{-0.8cm} dz  z^4 \, \bar{C}_{\gamma}[f](z) \bqty{1 - {\textstyle \frac{4\pi \tilde{w}_\text{max} \bar{p}_T^2}{z^2} }}^{\frac{1}{2}} \bqty{1 + {\textstyle \frac{4\pi \tilde{w}_\text{max} \bar{p}_T^2}{2 z^2 }}}\bigg). \nonumber
\end{align}
Importantly, in the limit of infinite expansion, i.e. $\wtilde_\text{max}\rightarrow\infty, \wtilde_\text{min}\rightarrow 0$, the spectrum is exactly a power law
\begin{align}
    \frac{dN_\gamma}{d^2\x_T d^2\p_T dy} &= \frac{\nu_\gamma}{(2\pi)^3} \frac{(\eta/s)^2}{\bar{p}_T^4}\tilde{C}_\gamma^\text{ideal}\label{eq:ideal}
\end{align}
where the normalization constant $\tilde{C}_\gamma^\text{ideal}$ is given by
\begin{align}\label{eq:CIdealDef}
    \tilde{C}_\gamma^\text{ideal} = 4 \int_0^\infty d\left(\frac{p}{T}\right) ~\left(\frac{p}{T}\right)^4 \Bar{C}^{\rm eq}_\gamma \left(\frac{p}{T}\right) \, .
\end{align}
For the three couplings we study, $\lambda=g^2N_C=5,10,20$, we obtain $C^\text{ideal}_\gamma=0.580,1.019,1.791$ and $\eta/s=2.716,0.994,0.385$ respectively.

\subsection{Pre-equilbirium spectrum}
\begin{figure}
\begin{center}
\centering
\begin{subfigure}[b]{0.45\textwidth}
\includegraphics[width=\textwidth]{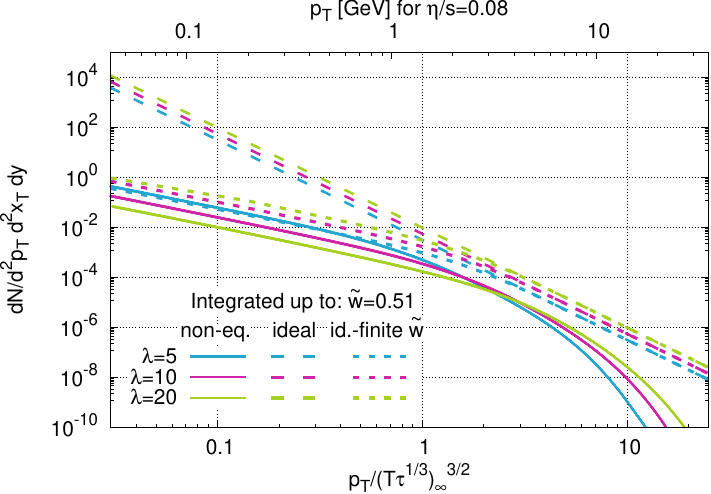}
\vspace{0.5pt}

\includegraphics[width=\textwidth]{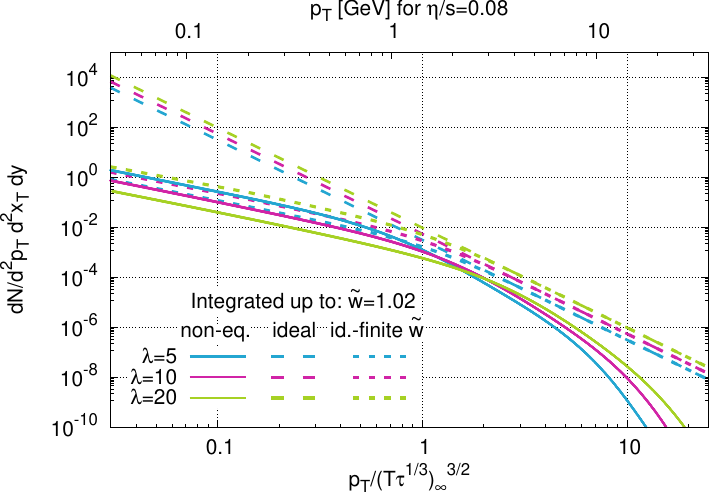}
\caption{Photon spectra}\label{fig:pTspectruma}
\end{subfigure}
\begin{subfigure}[b]{0.45\textwidth}
\includegraphics[width=0.89\textwidth]{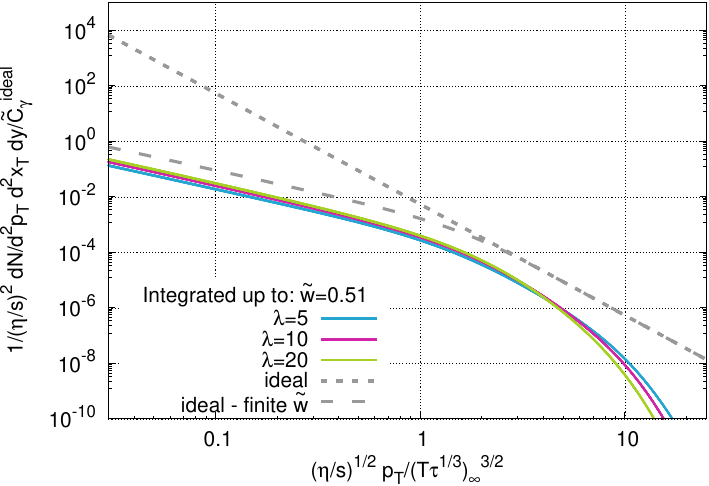}
\vspace{1.1cm}

\includegraphics[width=0.89\textwidth]{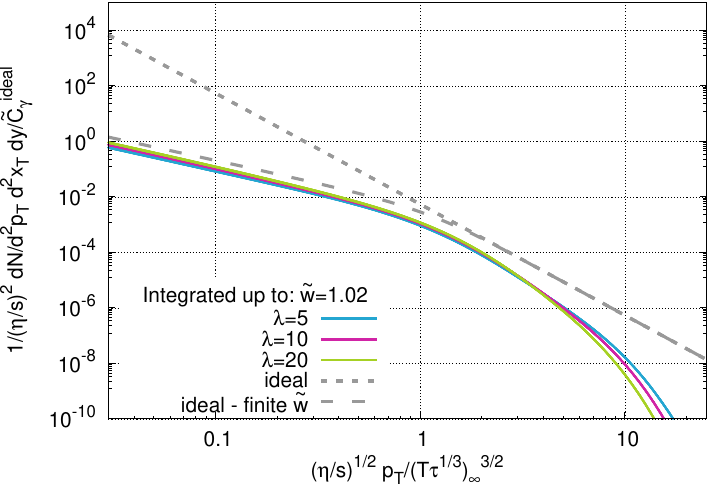}
\caption{Universal scaling curve}\label{fig:pTspectrumb}
\end{subfigure}
\caption{\label{fig:pTSpectrum} Spectrum of photons produced during the pre-equilibrium stage as a function of (a) $p_T/(T\tau^{1/3})_\infty^{3/2}$ and (b) $(\eta/s)^{1/2}p_T/(T\tau^{1/3})_\infty^{3/2}$. Different colors correspond to different coupling strengths $\lambda=5,10,20$ and different panels correspond to different final integration times $\tilde w=0.5,1.0$. The spectra in the right panels are divided by $(\eta/s)^2 ~\tilde{C}_\gamma^\text{ideal}$.}
\end{center}
\end{figure}

We present our results for the photon spectrum in \Cref{fig:pTSpectrum} for different couplings $\lambda=5,10,20$ and for different integration times $\wtilde=0.51,1.02$. We compare the non-equilibrium spectrum (solid) to the idealized Bjorken spectrum (dashed) (see \Cref{eq:idealfinite}), which is a simple power law, as well as to the idealized Bjorken spectrum with finite $\wtilde_\text{max}$, while $\wtilde_\text{min}=0$ (dotted) (see \Cref{eq:ideal}). The left panel shows the photon spectra $dN/d^2\p_Td^2\x_Tdy$, while the right one shows the universal scaling function $\mathcal{N}_{\gamma}(\tilde{w},\bar{\mathbf{p}}_T)$.
Looking at \Cref{fig:pTspectruma}, we see that in the thermal case a stronger coupling also means a higher production rate. However, for the non-equilibrium case for the same $\wtilde$ this is only true for high $p_T$ before the hierarchy switches and we find the reverse order in the soft regime. 
At early times the system is mainly gluon dominated while quarks get produced over time through gluon fusion ($gg\rightarrow q\bar q$) and splitting ($g\rightarrow q\bar q$). This is represented in the photon spectrum for high $p_T$. As this regime is produced at the earliest stages of the time evolution, quarks need to be produced before any photons can be produced. As a consequence of this, the pre-equilibrium photon production is overall suppressed compared to the thermal spectra.

When scaling the photon spectrum according to \Cref{eq:UniversalScaling} (see \Cref{fig:pTspectrumb}) the thermal spectra naturally collapse onto each other (gray dotted and dashed). However, also the EKT spectra (solid lines) agree for all three couplings, especially at late times. Only the high-$p_T$ sector does not scale according to $\mathcal{N}_\gamma$, as this sector is produced at early times and is barely modified afterwards. However, on this timescales an attractor description is not appropriate since the system is too far off equilibrium, which describes the differences between the couplings.

\section{Phenomenology}
\begin{wrapfigure}{r}{0.43\textwidth}
    \begin{center}
    \includegraphics[width=0.4\textwidth]{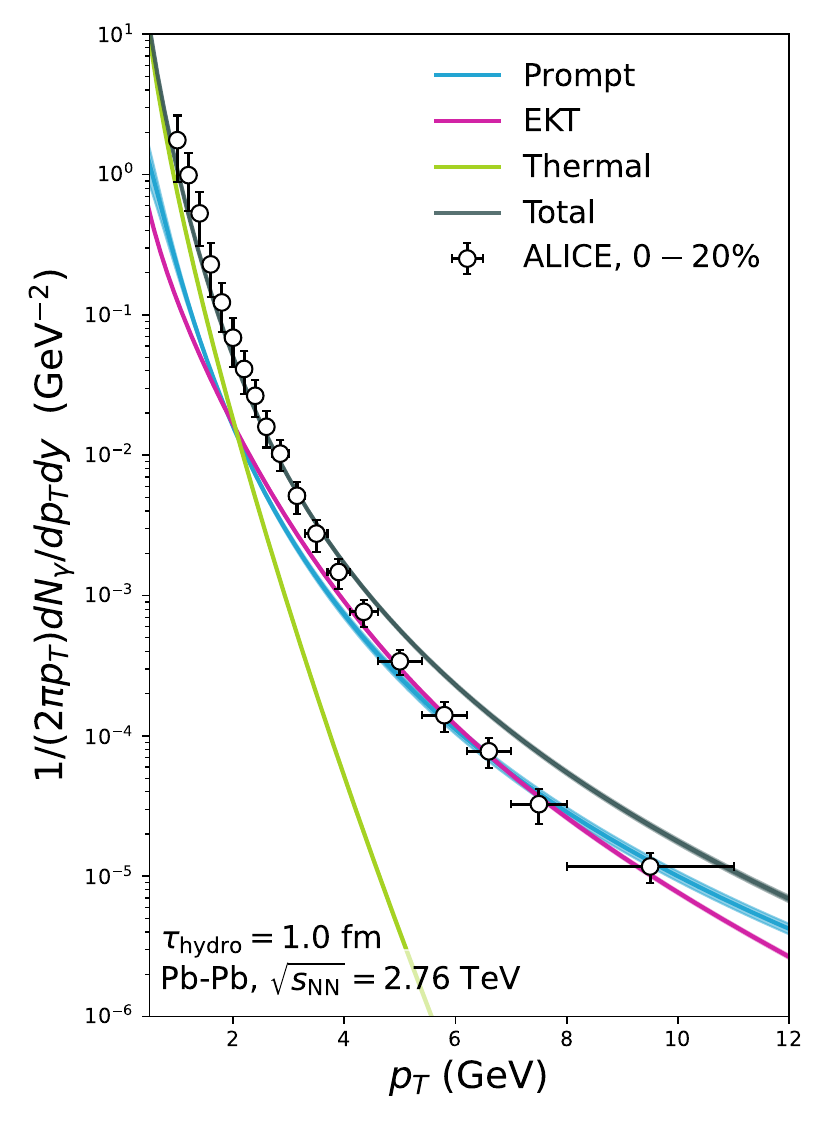}
    \end{center}
    \caption{\label{fig:pTYieldALICE} The differential photon spectra for $0-20\%$ centrality PbPb collisions at 2.76 TeV. ALICE results are shown as points. Solid red line shows the total computed photon spectra, which consists of prompt (blue), pre-equilibrium (green) and thermal rates in the hydrodynamic phase (yellow).}
\end{wrapfigure}

A remarkable feature of the scaling formula \Cref{eq:UniversalScaling} is that it allows us to compare our results to experimental data from high-energy heavy-in collisions. For this, the evolution in QCD kinetic theory is matched to the initial conditions of the following hydrodynamic evolution. Physically, this means that the temperature $T(\tauhydro,\x_T)$ in the fluid cell $\x_T$ at the initialization time of the hydrodynamic evolution is matched to the final state temperature of the preceding pre-equilibrium evolution. By using the temperature and initialization time, the scaled time variable $\wtilde(\x_T)$ for any given $\eta/s$ as well as $\left( \scriptstyle T(\x_T)\tau^{1/3} \right)_\infty^{3/2}$ can be extracted for each fluid cell $\x_T$. The scaling formula $\mathcal{N}_\gamma \left(\Tilde{w}(\x_T),\sqrt{\eta/s} ~p_T/\left( \scriptstyle T(\x_T)\tau^{1/3} \right)_\infty^{3/2} \right)$ then provides the pre-equilibrium photon spectrum produced by the cell $\x_T$. We should note that $\eta/s$ acts as an external paramter here and is not the $\eta/s$ extracted from the non-equilibrium evolution. Here we will use $\eta/s=0.08$. Further technical details on this implementation are given in \cite{Garcia-Montero:2023lrd}.

In \Cref{fig:pTYieldALICE} we show the comparison of prompt photons (blue), thermal photons from the QGP and the hadronic stage (green), pre-equilibrium photons (magenta) and total (gray) contributions to the photon yield as a function of $p_T$ compared to ALICE data \cite{ALICE:2015xmh} for Pb-Pb collisions at 2.76 TeV for the $0-20\%$ centrality class.

Comparing the different sources, that the thermal photon production is most pronounced at lower momentum levels, whereas EKT photons exhibit the least significant influence on photon spectra within this range. At approximately $p_T\sim 2.5$~ GeV, the thermal contribution becomes the smallest one and drops off rapidly with increasing transverse momentum. Between $p_T\sim 2.5$~ GeV and $p_T\sim 7$~ GeV the EKT photons are comparable to the spectra coming from prompt photons. Above $p_T\sim 7$~ GeV the EKT contributions decline and the hard-momentum sector is mainly dominated by prompt photons. In comparison to the ALICE data we overall find a good agreement.

\section{Summary and Outlook}
We computed the photon emission from the pre-equilibrium phase of the QGP using leading order QCD kinetic theory. We showed that photons with high $p_T$ are mainly produced at the earliest stages of the collision. Within this regime we observe a steep fall-off compared to the thermal rate due to a strong quark suppression at this stage of the evolution. Conversely, as time progresses, the soft regime is produced, where the photon spectra shows a characteristic power law behavior. 

Assuming that the pre-equilibrium evolution of the QGP can be described by a single scaled time variable $\tilde{w}$, we showed that the photon spectra satisfies a simple scaling formula where the momentum is scaled by ${(\eta/s)^{1/2}} \p_T/{(\tau^{1/3} T)^{3/2}_\infty}$ together with an additional overall normalization of $1/(\eta/s)^2\tilde{C}^\text{ideal}_\gamma$ to the photon spectra. This allows us to compare the pre-equilibrium contribution to different sources of photons during a heavy-ion collision. At soft momentum the photon yield is dominated by thermal photons from the QGP and the hadronic stage. However, above $p_T\sim2.5$GeV, the contributions from pre-equilibrium and prompt photons are larger than the thermal yield. EKT and prompt yields are almost of the same order although at soft and hard $p_T$ the EKT yield is notably smaller than the prompt one.

Since there is indication that dileptons from the pre-equilibrium phase might dominate the production in a certain range of invariant mass \cite{Coquet:2021gms,Coquet:2021lca}, a logical next step would be to include them into our calculations and find a similar scaling behavior as for photons. Due to exceeding other contributions in a certain mass range, they might be a good observable in order to gain direct information about the pre-equilibrium phase of high energy heavy-ion collisions.


\begin{thebibliography}{99}
\bibitem{Schlichting:2019abc}
S.~Schlichting and D.~Teaney,
Ann. Rev. Nucl. Part. Sci. \textbf{69} (2019), 447-476
doi:10.1146/annurev-nucl-101918-023825
[arXiv:1908.02113 [nucl-th]].


\bibitem{Berges:2020fwq}
J.~Berges, M.~P.~Heller, A.~Mazeliauskas and R.~Venugopalan,
Rev. Mod. Phys. \textbf{93} (2021) no.3, 035003
doi:10.1103/RevModPhys.93.035003
[arXiv:2005.12299 [hep-th]].


\bibitem{Arnold:2002zm}
P.~B.~Arnold, G.~D.~Moore and L.~G.~Yaffe,
JHEP \textbf{01} (2003), 030
doi:10.1088/1126-6708/2003/01/030
[arXiv:hep-ph/0209353 [hep-ph]].


\bibitem{Du:2020dvp}
X.~Du and S.~Schlichting,
Phys. Rev. D \textbf{104} (2021) no.5, 054011
doi:10.1103/PhysRevD.104.054011
[arXiv:2012.09079 [hep-ph]].


\bibitem{Coquet:2021gms}
M.~Coquet, X.~Du, J.~Y.~Ollitrault, S.~Schlichting and M.~Winn,
Nucl. Phys. A \textbf{1030} (2023), 122579
doi:10.1016/j.nuclphysa.2022.122579
[arXiv:2112.13876 [nucl-th]].


\bibitem{Coquet:2021lca}
M.~Coquet, X.~Du, J.~Y.~Ollitrault, S.~Schlichting and M.~Winn,
Phys. Lett. B \textbf{821} (2021), 136626
doi:10.1016/j.physletb.2021.136626
[arXiv:2104.07622 [nucl-th]].


\bibitem{Giacalone:2019ldn}
G.~Giacalone, A.~Mazeliauskas and S.~Schlichting,
Phys. Rev. Lett. \textbf{123} (2019) no.26, 262301
doi:10.1103/PhysRevLett.123.262301
[arXiv:1908.02866 [hep-ph]].


\bibitem{Garcia-Montero:2023lrd}
O.~Garcia-Montero, A.~Mazeliauskas, P.~Plaschke and S.~Schlichting,
[arXiv:2308.09747 [hep-ph]].


\bibitem{ALICE:2015xmh}
J.~Adam \textit{et al.} [ALICE],
Phys. Lett. B \textbf{754} (2016), 235-248
doi:10.1016/j.physletb.2016.01.020
[arXiv:1509.07324 [nucl-ex]].



\end{thebibliography}
\end{document}